# Chemical Environment Adaptive Learning for Optical Band Gap Prediction of Doped Graphitic Carbon Nitride Nanosheets

Chen Chen[1]†, Enze Xu[2]†, Defu Yang[1], Chenggang Yan[1], Tao Wei[3]*, Hanning Chen[4]*, Yong Wei[5]*, Minghan Chen[2]*

[1]Intelligent Information Processing Laboratory, Hangzhou Dianzi University, Hangzhou, China

[2]Department of Computer Science, Wake Forest University, Winston-Salem, NC, USA

[3]Department of Chemical Engineering, Howard University, Washington, D.C., USA

[4]Texas Advanced Computing Center, the University of Texas at Austin, Austin, TX, USA

[5]Department of Computer Science, High Point University, High Point, NC, USA

†Equal contribution; *Corresponding author

Lead contact email: chenm@wfu.edu

**Abstract**

This study presents a new machine learning algorithm, named Chemical Environment Graph Neural Network (ChemGNN), designed to accelerate materials property prediction and advance new materials discovery. Graphitic carbon nitride (g-$C_3N_4$) and its doped variants have gained significant interest for their potential as optical materials. Accurate prediction of their band gaps is crucial for practical applications; however, traditional quantum simulation methods are computationally expensive and challenging to explore the vast space of possible doped molecular structures. The proposed ChemGNN leverages the learning ability of current graph neural networks (GNNs) to satisfactorily capture the characteristics of atoms' chemical environment underlying complex molecular structures. Our experimental results demonstrate more than 100% improvement in band gap prediction accuracy over existing GNNs on g-$C_3N_4$. Furthermore, the general ChemGNN model can precisely foresee band gaps of various doped g-$C_3N_4$ structures, making it a valuable tool for performing high-throughput prediction in materials design and development.

**Teaser**

This study introduces a new graph neural network to predict band gaps in graphitic carbon nitride with high accuracy.

**Introduction**

Graphitic carbon nitride (g-$C_3N_4$) is one of the oldest synthetic polymers reported in 1834 by Berzelius and Liebig (*1*). Due to its chemical inertness and insolubility in most common solvents (*2*), g-$C_3N_4$ was rarely explored until 2009, when it was used as a photocatalyst for hydrogen production through water-splitting (*3*). The extraordinary photocatalytic performance of g-$C_3N_4$ is primarily ascribed to its band gap of 2.7 eV (*4*), which places its conduction band edge above the proton reduction potential, and its valence band edge below the water oxidation potential. Inspired by this pioneering study, many efforts have been carried out to engineer the g-$C_3N_4$'s band gap for various photoelectrochemical applications, such as dye-sensitized solar cells (*5*), biomedical sensors (*6*), photodynamic cancer therapy (*7*), photothermally accelerated wound healing (*8*), and water purification (*9*). For example, a g-$C_3N_4$ nanosheet co-doped by sulfur and boron reduces the band gap to 2.5 eV, resulting in a more efficient visible-light-driven hydrogen production because of a better match between the co-doped g-$C_3N_4$'s absorption spectrum and the solar power spectrum (*10*). In a study of water purification (*9*), the band gap of a g-$C_3N_4$ nanosheet was found to be drastically reduced to 1.9 eV upon the substation of a carbon atom by a phosphorous atom (*9*). More interestingly, it was discovered that the band gap not only depends on a dopant's element type but also is subject to its substitution site. For instance, the substitution by nitrogen at two chemically inequivalent carbon sites yields distinctive band gaps of 2.57 eV and 2.90 eV (*9*), suggesting the importance of atomically precise doping for desired photophysical properties of a photocatalyst such as doped g-$C_3N_4$. With the prosperous and rapid progress of single-atom catalysts over the past decade (*11*), the precise tuning of band gaps through doping has become feasible, paving the way for a systematic exploration of the optimal doping scheme for a given photochemical function of g-$C_3N_4$.

Due to the importance of photophysical properties in the discovery and design of g-$C_3N_4$, it is critical to accurately predict them utilizing molecular structures and atomic characteristics, as has been widely done in other functional materials. For instance, a group of emergent high-temperature superconductors was discovered in doped ferroelectrics due to a remarkable electron-phonon coupling when the dopants move the ferroelectrics' Fermi surface above their conduction band edge (*12*). Traditional methods use quantum simulations to estimate the band gap, such as *ab initio* molecular dynamics (AIMD) (*13, 14*), quantum Monte Carlo (*15, 16*), and density functional theory (*17, 18*). However, these usually are computationally expensive, particularly for complex systems. Recently, large-scale quantum chemical calculations, molecular dynamics simulations, and high-throughput experiments have produced unprecedented amounts of data for analyses. When machine learning is applied to material property prediction, it provides an efficient and convenient way of predicting promising molecules from a pool of candidates (*19*) and even proposing novel molecules (*20*) through a systematic exploration of structure-property relationships in chemical space (*21, 22*). For example, heptazine, the building block of g-$C_3N_4$, consists of five chemically unique doping sites. Even if we consider only twenty elements as potential dopants for a triply doped g-$C_3N_4$, there are millions of possible chemical structures, which makes quantum simulation a daunting task that can only be resolved by machine learning.

To predict the optical band gaps of graphitic carbon nitride and its doped variants using their molecular structures, the very first yet challenging step is to form a permutation-invariant representation of the three-dimensional non-Euclidean molecular structures. In the context of machine learning, graphs have been used to fulfill the need to represent molecule structures (*23*), where atoms are treated as nodes and chemical bonds as edges. Graph neural networks (GNNs), such as graph convolutional network (GCN) (*24*) and message passing neural networks (MPNN) (*25*), are popular deep learning models designed specifically to learn the graph representation for downstream prediction tasks, including classification and regression. GCN has found extensive application in the data analyses of molecular dynamics (*26*) and medical diseases (*27, 28*), and MPNN generalizes multiple categories of spatial GCNs to learn molecular features (*29-31*) and shows promising results in molecular property prediction (*32-36*). Both methods rely on aggregating messages from a node and its neighbors to generate the node's representation. In MPNN, messages are passed among neighboring nodes via a message function, the node embeddings are then updated through a vertex function, and the resulting molecule feature representations are generated using a readout function from the node embeddings in the graph.

The conceptual similarity between GNNs and the underlying chemical bonding topology has led to the widespread use of GNNs in predicting quantum mechanical properties. For instance, in a crystal graph convolutional neural network (CGCNN) (*37*) study, eight physical properties, such as band gap and Fermi energy, were accurately predicted for 46744 crystals. In another equivariant message passing neural network study (*38*), rotationally equivariant representation was proposed to enable the prediction of tensorial properties and molecular spectra. Graph isomorphism network (GIN) (*39*) is designed for discrete feature space to distinguish isomorphic graph structures in practical applications. More recently, the neural equivariant interatomic potentials (NequIP) (*40*) was developed to perform molecular dynamics simulations at the accuracy of density functional theory while demanding a fractional of its computational cost. These studies showcase the efficacy of GNNs in capturing complex relationships and spatial arrangements, enhancing our understanding of chemical systems and accelerating discovery in materials science and quantum chemistry.

While the aforementioned GNN models use a single aggregation function ($Sum$ or $Mean$) to generate node features, this may not effectively exploit an atom's chemical environment, which affords different band gaps in various molecules through diverse interatomic interactions. As a result, such node features are insufficient to identify local molecular structures. As illustrated in Fig. 1, using a single aggregation function fails to differentiate neighborhood messages from different molecular structures (detailed explanations are given in the Methods section). It can lead to poor node representations that are unable to reflect the local chemical environment characteristics, and thus inaccurate structure-dependent property prediction, such as band gaps.

To address this challenge, we proposed a new Chemical Environment Graph Neural Network (ChemGNN) that utilizes chemical environment adaptive learning (CEAL) layers to effectively extract deep information from the neighboring environment of each node. The ChemGNN model can automatically adapt multiple aggregators to provide valuable insights into the chemical environments of atoms to generate representations of nodes. Previous works in the field have also explored the use of multiple aggregation functions for graph data learning. GraphSAGE (*41*) utilizes multiple aggregators to combine node features from a fixed-size sampled neighborhood. However, in GraphSAGE, no learnable weights are assigned specifically to the aggregated messages, which are then used individually in a sequence to update the features of the central node. The order of aggregators may impact the model performance. In ChemGNN, we leverage aggregated neighborhood messages as a collective input to the MLP for feature updating. Therefore, the order of aggregators is not a concern in our approach. Inspired by cognitive attention, graph attention network (GAT) (*42*) implicitly assigns different weights to nodes in a neighborhood based on their messages.

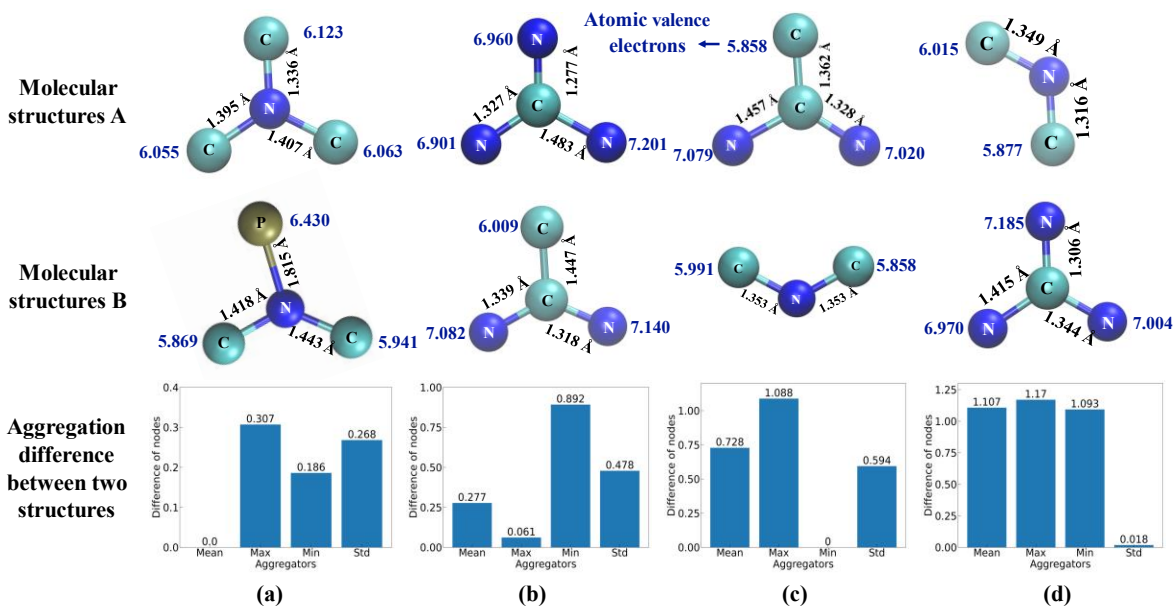

**Fig. 1. Four examples demonstrate that a single aggregator cannot distinguish between different g-C$_3$N$_4$ structures.** Four aggregation functions ($Mean$, $Max$, $Min$, and $Std$) are employed to aggregate messages (atomic valence electrons) from neighboring nodes for the central atom. The bar plots illustrate the differences in values among the four aggregations of molecule structure pairs (A and B) in a column, respectively. It is shown that using a single aggregator, $Mean$ in **(a),** $Max$ in **(b)**, $Min$ in **(c)**, and $Std$ in **(d)** alone fails to differentiate the molecule structure pairs in (a)-(d), respectively.

Recently, principal neighborhood aggregation (PNA) (*43*) introduced different scalers for the aggregated messages of central nodes, but these scalers are based on node degrees and are not learnable. It fails to capture the characteristics of atomic interactions when chemical composition changes. For example, consider the scenario where an atom is substituted with a different type of atom. While the node degree does not change, the characteristics of atomic interactions do change. This indicates that PNA's information acquisition of the chemical environment is not sufficiently comprehensive. Our approach, ChemGNN, addresses this limitation by employing learnable weights applied to the aggregation functions, enabling us to explore a more accurate representation of the complex interactions within chemical systems. Another related approach, Deepergcn (*44*), proposed a generalized aggregation function that is parameterized by a continuous variable. By learning the variable, the generalized aggregation function can be considered as a combination of $Mean$ and $Max$. However, it is unclear how to generalize the aggregation function to include statistical aggregations such as $Var$ and $Std$. Isotropic GNN (EGC-M) (*45*) employs multiple basis weights and node-wise weights for multiple aggregators. While this approach is close to our work, the use of multiple basis weights to aggregate messages can result in an excessive number of parameters when dealing with a large number of atoms in chemical systems.

Our studies show that the proposed ChemGNN algorithm can predict the optical band gaps of g-C$_3$N$_4$ nanosheets and eight of its doped variants more than 100% accurately than other GNN models, including GCN, GraphSAGE, GAT, MPNN, and PNA. More significantly, since the CEAL layers can effectively extract an atom's chemical environment characteristics, the ChemGNN models are expected to afford high prediction accuracy for other molecule properties, which largely depend on local interatomic interactions.

**Results**

Details of the ChemGNN algorithm are given in the Methods section. Concisely, ChemGNN is designed with CEAL layers (adaptive aggregation mechanism) to improve the information extracting ability. The architecture of a CEAL layer is illustrated in Fig. 2. A set of aggregators (detailed in Table 2) are exploited to collect various aspects of the chemical environment attributes of a node. The weights are adaptively learned in the training process to determine the optimal combination of aggregation functions based on the chemical environments of atoms. These chemical environment features are assigned with adaptively learnable weights to reflect their importance in determining the central node's representation. Compared with other GNN models, ChemGNN can gain effective insight into the local chemical environments to facilitate molecular property prediction. To demonstrate the advantage of the proposed model, $g-C_3N_4$ and its eight doped variants are selected as the research objects to predict their optical band gaps.

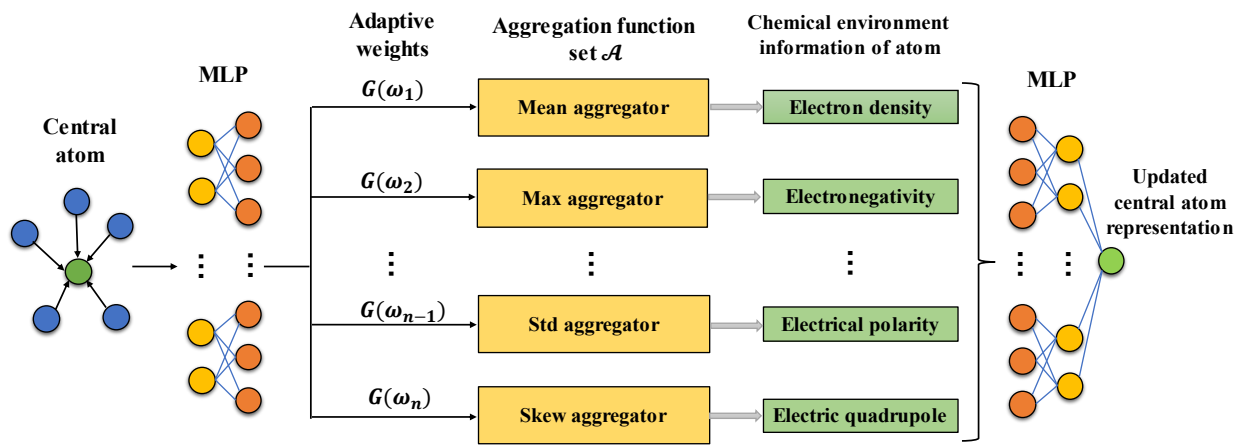

**Fig. 2. The architecture of a CEAL layer.** A set of aggregators are utilized to extract various attributes of the chemical environment of an atom. Adaptively learnable weights are assigned to the aggregators to reflect their importance in determining the central node's representation. MLPs (multilayer perceptron) are applied as pre/post-processing layers to enhance the expressiveness of a CEAL layer.

The experiments were conducted using PyTorch Geometric, implemented on an NVIDIA RTX 3090 Ti GPU with 24GB RAM. A global sum pooling was employed as the readout layer for all models. The training process for all models utilized the Adam optimizer (*46*) with a plateau learning rate scheduler. The initial learning rate was set to 0.001, with a drop factor of 0.5. The patience for the learning rate scheduler was set to 30 epochs, and a lower bound of 0.0001 was imposed on the learning rate. The batch size is 64~128, and the maximum number of epochs is 400. To avoid overfitting, early stopping is utilized with a patience of 30 epochs. Each dataset is randomly partitioned into the training (60%), validation (10%), and testing (30%) sets, respectively. To ensure the robustness of the results, all reported findings in this section were based on five independent experiments. All methods utilize the same features of nodes and edges for comparison, but the processing methods for these features vary from model to model.

Water and Heptazine

Aiming to benchmark our proposed ChemGNN model against GCN, we first compare their performance on the band gaps of water and heptazine, which are the most common molecule and the building block of g-$C_3N_4$, respectively. Our approach shows its advantages over GCN as molecular structures become slightly complex from water to heptazine. Water is the most common solvent of g-$C_3N_4$, and it only consists of one oxygen atom and two hydrogen atoms (Fig. 3(a)). For such a simple molecule, excellent performance is expected from any established machine learning algorithm due to the low complexity and high symmetry of water's molecular structure. In Fig. 3(b) and 3(c), nearly all data points fall very close to the diagonal line, illustrating a remarkable agreement between the predicted and true values obtained by both GCN and ChemGNN models. The average predicted optical band gap of ~10.8 eV is well in line with water's famous set of narrow bands near 115 nm, corresponding to a Rydberg transition at 10.7 eV (*47*).

Unlike the three-atom $H_2O$, heptazine has six oxygen, seven nitrogen, and three hydrogen atoms (Fig. 3(d)). Due to its notably increased degrees of freedom, a more advanced machine learning algorithm is required to accurately map its molecular structures to its optical band gaps. As presented in Fig. 3(e) and 3(f), GCN exhibits a cluster pattern, whereas ChemGNN produces a dense linear regression pattern. This distinction highlights the superior performance of ChemGNN compared to GCN. More specifically, our model has a much higher fitting coefficient ($R^2$) value than GCN (0.213 vs. 0.913). Our model also yields a lower mean absolute error (MAE) than GCN (0.031 eV vs. 0.071 eV). The average predicted optical band gap of ~3.8 eV for heptazine is consistent with the experimental value of 3.7 eV (*48*).

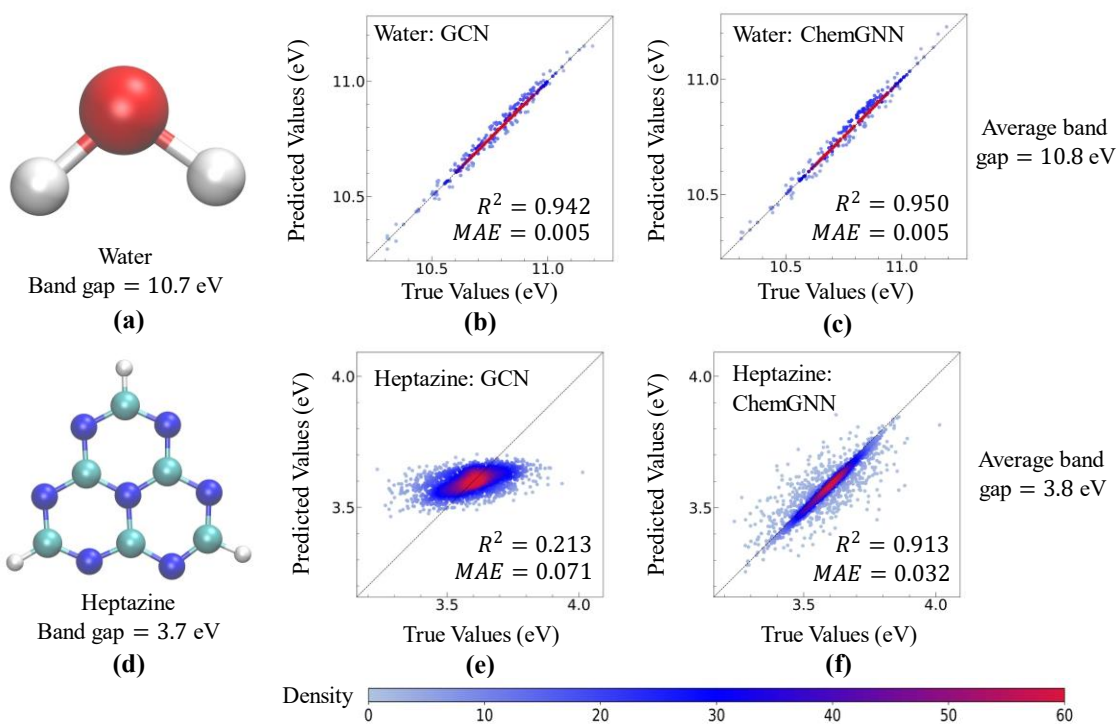

**Fig. 3. Predicted vs. true band gaps of water and heptazine molecules using GCN and ChemGNN. (a)** Molecular structure of water. The fitting coefficient $R^2$ and MAE indicate that simple water molecules can be almost equally well handled by both GCN **(b)** and ChemGNN **(c)**. **(d)** Molecular structure of heptazine. The ChemGNN model **(f)** significantly outperforms GCN **(e)** in predicting the optical band gaps of slightly more complex heptazine.

**g-C3N4 and its Doped Variants**

As shown in Fig. 4(a), a pristine g-$C_3N_4$ nanosheet consists of heptazine units connected by tertiary nitrogen atoms. For each heptazine unit, there are three chemically inequivalent nitrogen sites, ($N_1$, $N_2$, and $N_3$ in Fig. 4(b)), while the number of chemically inequivalent carbon sites is two ($C_1$ and $C_2$ in Fig. 4(b)). In the present study, we explored the substitution of nitrogen for carbon and phosphorous, in addition to the doping of phosphorous at the carbon sites. Each doped compound is labeled as A→B, where A is the doping site and B is the dopant's element. For instance, $N_1$→P refers to the doped g-$C_3N_4$ wherein the nitrogen atom at an $N_1$ site is substituted by a phosphorous atom. As a result, nine g-$C_3N_4$ compounds, including the undoped ones, were investigated. Specifically, each compound is represented by a 3×3 supercell (Fig. 4(a)) with an experimental crystal structure of a = 7.13Å, b = 7.13Å, and γ = 60° determined by X-ray diffraction (*49*). Since the band gap of a g-$C_3N_4$ nanosheet is sensitive to its atomistic structure, which notably changes upon thermal fluctuation, a large ensemble of atomistic configurations obtained from our quantum-based AIMD simulations is needed to fully understand the structure-band gap relationship by accounting for the thermal effect, especially at room temperature. Therefore, approximately 10,000 atomistic configurations were randomly extracted for each compound from a 1-ns AIMD trajectory.

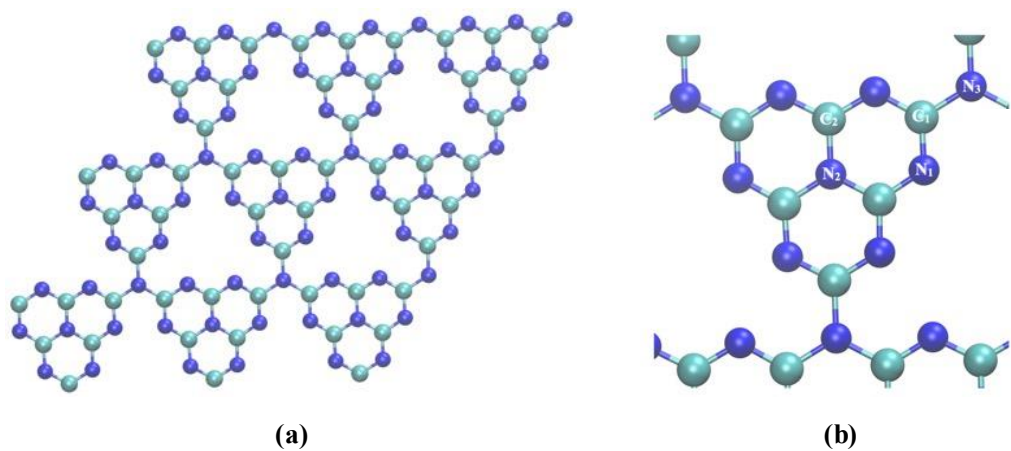

(a)          (b)

**Fig. 4. Molecular structure of g-C3N4 and its doping sites.** **(a)** Molecular structure of a 3×3 supercell of g-$C_3N_4$. The carbon and nitrogen atoms are colored cyan and blue, respectively. **(b)** Designated doping sites (i.e., $N_1$, $N_2$, $N_3$, $C_1$, and $C_2$) in the heptazine unit of an undoped g-$C_3N_4$ nanosheet.

Nine datasets of approximately 110,000 atomistic configurations and their optical band gaps of the pristine g-$C_3N_4$ nanosheet and its eight doped variants were obtained by the AIMD simulations. Those datasets (Undoped, C1P, C2P, N1C, N2C, N3C, N1P, N2P, and N3P, see detail in subsection g-$C_3N_4$ and Its Doped Variants) have 22237, 10165, 10715, 10305, 13217, 10570, 13809, 12981, and 11781 atomistic configurations, respectively. For the odd-electron datasets (C1P, C2P, N1C, N2C, and N3C), the band gaps for both the alpha and beta spin channels were calculated due to the broken spin symmetry. By contrast, for the even-electron datasets (Undoped, N1P, N2P, and N3P), only the band gaps for the alpha spin channel were evaluated due to their spin-paired orbitals.

We applied our ChemGNN model to predict the band gaps of g-$C_3N_4$ and its doped variants. A ChemGNN model with 4 CEAL layers (see the Discussion section for the selection of layer number) was used to perform optical band gap prediction. The following chemical characteristics were used as the initial node embedding, i.e., coordinates of atoms in space, atom type, and electron numbers on the 1s, 2s, 2p, 3s, 3p, and 3d angular momentum channels. For detailed information on model settings and fine-tuned parameters for each model, please refer to supplementary Table S1.

As a first step, to illustrate the predictive power of our proposed approach on a single category of g-$C_3N_4$ molecular structures, the ChemGNN model was trained and tested on the N1C dataset. The dataset contains 10,305 molecular structures and their optical band gaps of the alpha and beta spin channels obtained by AIMD simulations. Figures 5(a) and 5(b) illustrate the accurate prediction capability of our proposed ChemGNN model for the optical band gaps of the alpha (red) and beta (blue) spin channels of N1C. In comparison to GCN with an MAE of 0.123 eV, our ChemGNN model reduces the MAE to 0.035 eV, demonstrating remarkable performance enhancement. In Fig. 5(c), the predicted distribution of optical band gaps aligns nearly perfectly with that of true distribution even for odd-electron systems, such as N1C, featuring distinct band gaps for alpha and beta spin channels.

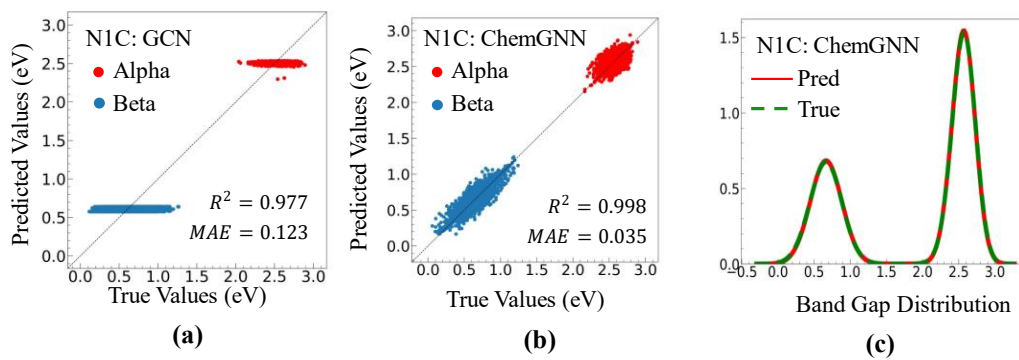

**Fig. 5. Performance comparison between GCN and ChemGNN on the N1C dataset.** Predicted vs. true band gaps of the N1C dataset with the Alpha (red) and Beta (blue) spin channels, using the GCN **(a)** and ChemGNN **(b)**, respectively. **(c)** Distributions of the predicted (red solid line) vs. true optical band gaps (green dashed line) of N1C.

Secondly, to demonstrate the superior capabilities of ChemGNN in accurately predicting the optical band gaps of molecules with diverse categorical structures, our model was trained using all nine datasets containing g-$C_3N_4$ and its eight doped variants. After the model was trained, testing data containing a mixture of the above nine categories of structures were fed into the model and their optical band gaps were predicted. The performance of the proposed model was compared with those of other established GNN models, including GCN, GAT, GraphSAGE, MPNN, and PNA. Fig. 6(a) shows that the ChemGNN model has the lowest MAE (0.031 eV), indicating a significant improvement in prediction accuracy. The average MAE of the predicted optical band gaps attained by our ChemGNN over five experiments is 106%, 84%, 116%, 100%, and 38% lower respectively than those of GCN, GraphSAGE, GAT, MPNN, and PNA. Furthermore, as shown in Fig. 6(b-g), the proposed model yields optical band gap predictions much more closely concentrated along the diagonal line, indicating the superiority of ChemGNN over other established GNN models. All these results strongly affirm our model's salient capabilities to accurately extract information from the chemical environment of atoms for the purpose of mapping molecular structures to optical band gaps. Importantly, this ability to effectively extract the local chemical environment information would enable the accurate prediction of any other molecular properties that are subject to atomic interactions, accelerating the experimental discovery of novel materials with desired functions.

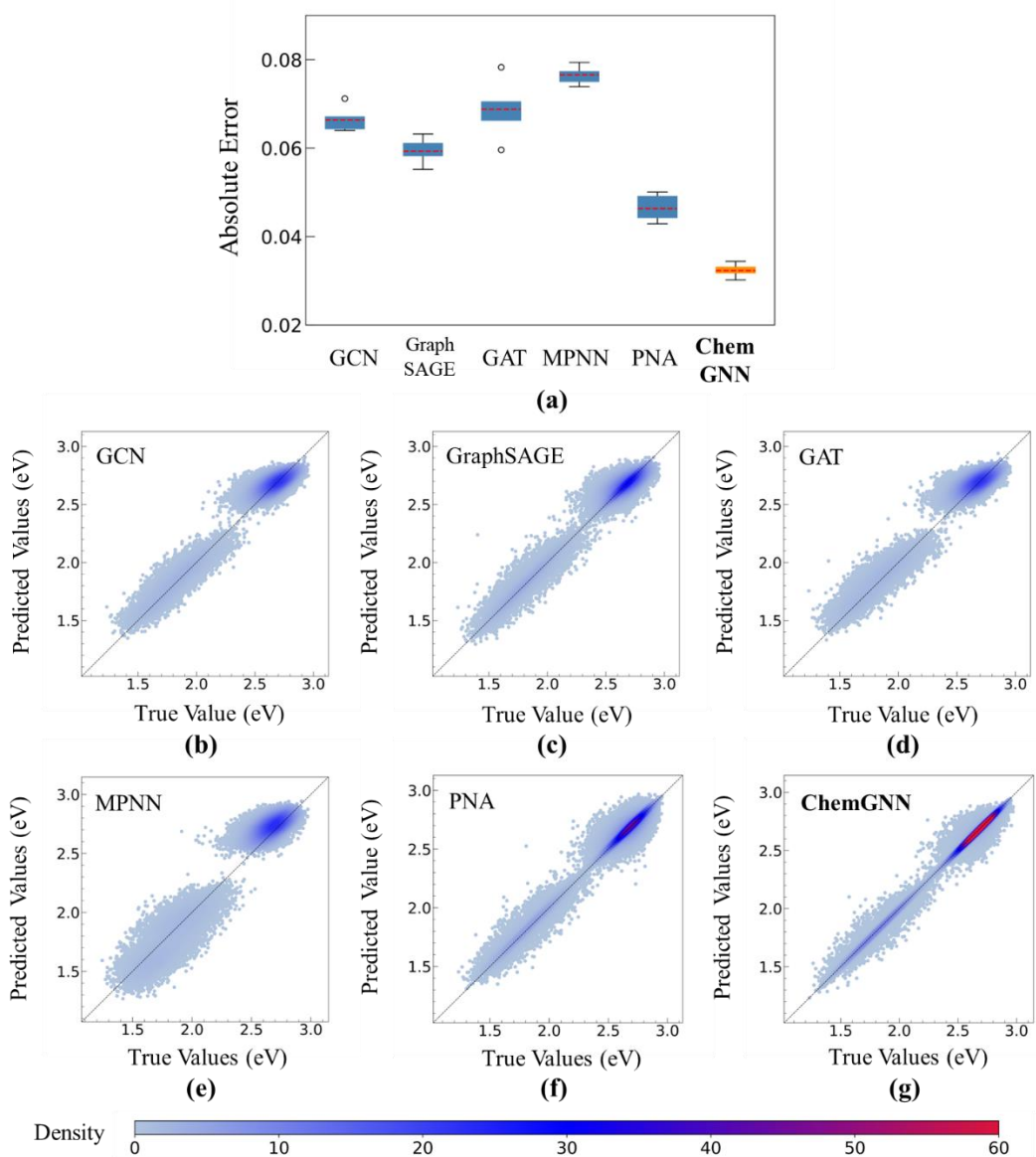

**Fig. 6. Performance comparison between GNN models and our ChemGNN. (a)** The boxplot of MAE under five independent experiments using ChemGNN and other established GNN models. **(b-g)** Predicted vs. true band gaps obtained by various GNN models. The densities of predicted band gaps are visualized by colors. All models were trained and tested using a mixture of data from g-$C_3N_4$ and its eight doped variants.

**Discussion**

Experimental results demonstrate that the proposed ChemGNN model consistently exhibited superior performance with respect to mean absolute error and $R^2$ of optical band gap predictions for molecule structures ranging from simple (such as water and heptazine) to complex (g-$C_3N_4$ and its eight doped variants). To further explore the category-wise performance of optical band gap prediction, our trained model used in Fig. 6 was tested on single-category data. Fig. 7 displays category-wise MAEs and predicted band gaps. Most of the predicted band gaps are prevalent along the diagonal line, evidenced by the high density of predicted data points. The category-wise performance of the proposed model with other established GNNs are listed in Table 1. The proposed model constantly yields significantly lower prediction errors than other GNN models. We also performed ablation studies to examine the isolated effect of the adaptive aggregation mechanism, see Fig. S1 in the supplementary. Moreover, to validate the effectiveness and applicability of our model, we evaluate its performance on two well-known datasets: QM9 and FePt, see Fig. S2 in the supplementary. Fine-tuned model hyperparameters are listed in Table S1 of the supplementary.

In conclusion, the proposed ChemGNN model uses the adaptive aggregation mechanism to extract deep insight from atoms' chemical environment, addressing the limits of using single aggregation. Experimental results show that the proposed model can significantly improve the optical band gap prediction of graphitic carbon nitride nanosheets and the doped variants. Moreover, the proposed model's learning power is promising, and it can be potentially applied to predicting other structure-dependent molecular properties, such as nuclear magnetic resonance chemical shifts. Overall, the ChemGNN model offers a promising approach to enhance predictions of molecular properties, which could have broad applications in various fields, including material science, drug discovery, and computational chemistry.

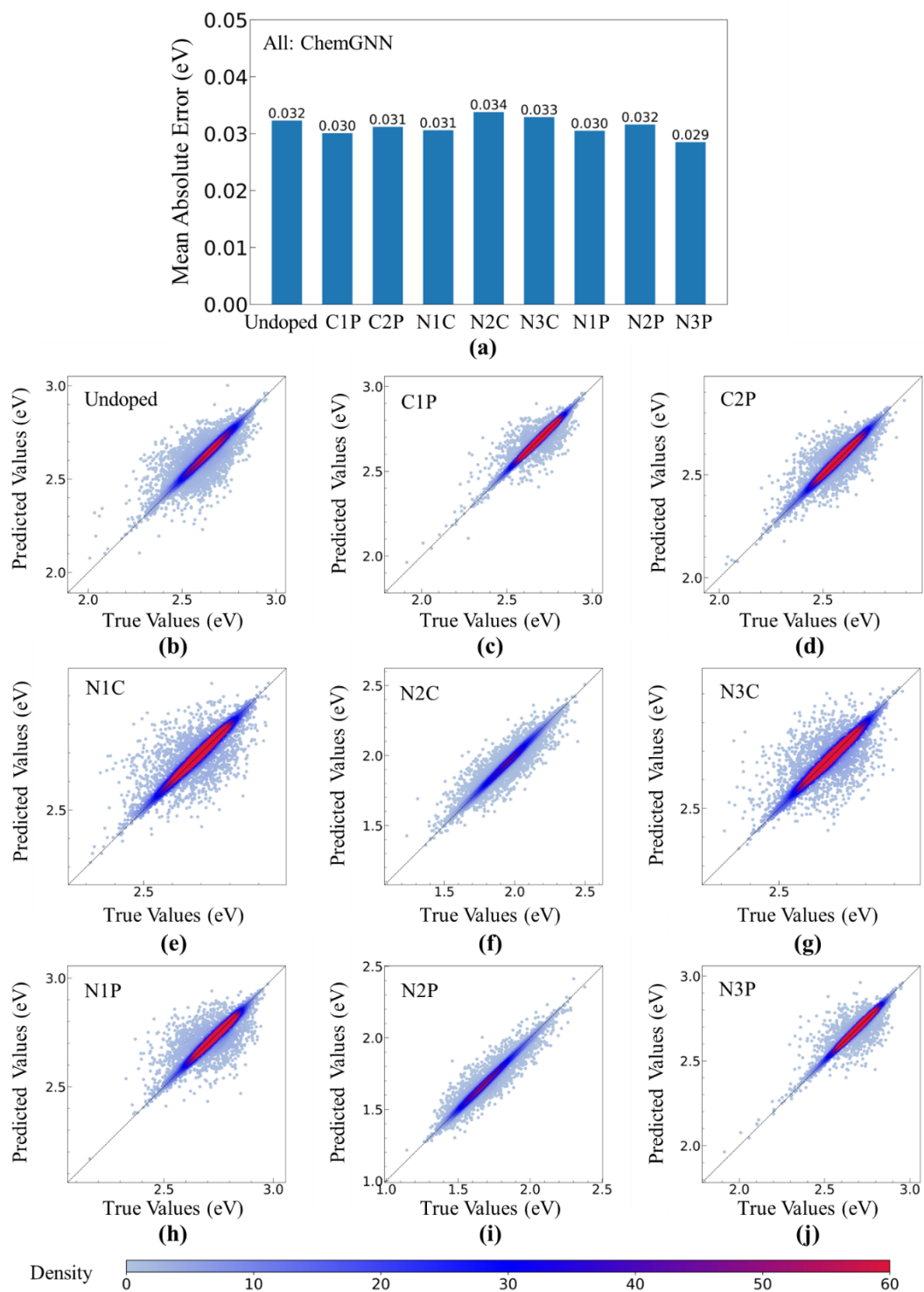

**Fig. 7. Category-wise performance of our ChemGNN.** The model was the same one trained in Fig. 6 but tested on category-wise data, including g-$C_3N_4$ and its eight doped variants. **(a)** Category-wise MAEs. **(b-j)** Predicted vs. true band gaps.

**Table 1. Category-wise MAEs (eV) obtained by the ChemGNN and other established GNN models.** The category-wise MAEs and standard deviations are averaged over five experiments.

| Models/Variants | GCN | GraphSAGE | GAT | MPNN | PNA | ChemGNN |
|---|---|---|---|---|---|---|
| Undoped | 0.0679±0.02 | 0.0537±0.03 | 0.0644±0.01 | 0.0680±0.02 | 0.0470±0.01 | **0.0324**±0.01 |
| C1P | 0.0662±0.01 | 0.0688±0.01 | 0.0719±0.03 | 0.0614±0.03 | 0.0463±0.03 | **0.0304**±0.02 |
| C2P | 0.0661±0.01 | 0.0676±0.01 | 0.0660±0.03 | 0.0761±0.02 | 0.0448±0.02 | **0.0315**±0.02 |
| N1C | 0.0732±0.02 | 0.0530±0.02 | 0.0475±0.02 | 0.0568±0.01 | 0.0486±0.02 | **0.0312**±0.01 |
| N2C | 0.0739±0.03 | 0.0498±0.04 | 0.0663±0.01 | 0.0632±0.01 | 0.0477±0.02 | **0.0344**±0.01 |
| N3C | 0.0648±0.01 | 0.0638±0.01 | 0.0727±0.04 | 0.0611±0.02 | 0.0518±0.02 | **0.0330**±0.01 |
| N1P | 0.0633±0.02 | 0.0582±0.02 | 0.0675±0.05 | 0.0597±0.02 | 0.0410±0.02 | **0.0303**±0.01 |
| N2P | 0.0621±0.02 | 0.0554±0.02 | 0.0753±0.02 | 0.0665±0.03 | 0.0507±0.01 | **0.0315**±0.02 |
| N3P | 0.0631±0.02 | 0.0599±0.03 | 0.0626±0.02 | 0.0595±0.03 | 0.0402±0.01 | **0.0285**±0.02 |
| Average | 0.0663±0.02 | 0.0589±0.02 | 0.0691±0.01 | 0.0634±0.01 | 0.0465±0.01 | **0.0315**±0.01 |

## Materials and Methods

### Quantum Mechanics Simulation for the Band Gap of Doped g-C3N4

Unless otherwise specified, all *ab initio* simulations in the present study were performed by the open-source CP2K software (*50*) with Goedecker-Teter-Hutter (GTH) pseudopotential (*51*), Heyd-Scuseria-Ernzerhof (HSE06) exchange-correlation functional (*52*), and polarized-valence-double-$\zeta$ (PVDZ) basis set (*53*). Our chosen HSE06 range-separated hybrid functional is justified by our calculated band gap of 2.78 eV for the optimized undoped g-$C_3N_4$ nanosheet. Moreover, for each of our selected atomistic configurations, its angular-momentum-resolved Mulliken charges (*54*) were also calculated for their direct relevance to the electron distribution that governs the band gap. As a result, the atomic coordinates and angular-momentum-resolved Mulliken charges were used as the input data for our machine-learning model to predict the band gaps of the undoped and doped g-$C_3N_4$ nanosheets.

### Graph Neural Networks (GNNs)

A graph $G$ consists of a vertex (also called a node) set $V$ and an edge set $\mathcal{E}$, i.e., $G = (V, \mathcal{E})$. The topology of a graph is described by the adjacency matrix $D$, which is a square matrix of size $M \times M$ in which $M$ is the number of nodes. $D(u, v) = 1$ if node $u$ is connected to node $v$, otherwise $D(u, v) = 0$. A node is represented by a feature (embedding) vector $X \in \mathbb{R}^{m \times 1}$ containing the embedded node properties, in which $m$ is the number of properties. Graphs are suitable for representing molecule structures because they are permutation invariant, i.e., when a molecule rotates in 3D space, the graph representing its structure does not change. When a graph is used to represent the structure of a molecule, an atom is represented as a node and a chemical bond between atoms is represented as an edge. Atomic characteristics are used as a node's initial embedding. GNN is a framework that computationally learns how to propagate information across the graph to compute node embeddings for downstream tasks such as optical band gap prediction. A GNN model consists of multiple layers. A GNN layer is a two-step computation process, including message computing and message aggregating. The two steps are defined as follows in equations (1) and (2).

$$m_u^{(k+1)} = msg^{(k+1)}\left(X_u^{(k)}\right), u \in \{N(v) \cup v\} \tag{1}$$

$$X_v^{(k+1)} = agg^{(k+1)}(\{m_u^{(k+1)}, u \in \{N(v) \cup v\}\}) \tag{2}$$

In Eq. (1), $N(v)$ is the set of neighboring nodes of node $v$, $X_u^{(k)}$ is the node feature at the $kth$ GNN layer. $msg^{(k+1)}$ and $m_u^{(k+1)}$ are the message function and message of node $u$ at the $(k+1)th$ layer, respectively. In Eq. (2), $agg^{(k+1)}$ and $X_v^{(k+1)}$ are the aggregation function and feature vector of node $v$ at the $(k+1)th$ layer. Different instantiations of GNN utilize various message and aggregation functions. Compared to established GNN models, such as GCN, GAT, GraphSAGE, and MPNN, the proposed ChemGNN model can effectively learn the deep features of nodes from graphs to significantly improve the structure-dependent property predictions, such as the optical band gaps of g-$C_3N_4$ and its doped variants.

**Chemical Environment Graph Neural Network (ChemGNN)**
As discussed in the Introduction section, Fig. 1 uses the atomic valence electrons as the sole node feature to illustrate the difference |Δ| between these aggregators for four distinct scenarios. In Fig. 1(a), the primary difference between structures A and B is the substitution of a carbon atom for a phosphorous atom. The more electropositive phosphorous atom loses approximately one electron to its neighboring nitrogen atoms, eventually carrying $a + 1$ net charge. Since a neutral phosphorous atom carries one more valence electron than its carbon counterpart, the $N_1{\rightarrow}P$ doping happens to preserve the mean valence electrons of the three atoms connected to the central nitrogen atom, resulting in a nearly negligible $|\Delta_{Min}|$ of 0.0002. By contrast, the N-P bond polarity is substantially greater than the N-C bond, affording a large $|\Delta_{Max}|$ of 0.3069 due to more electrons accumulated at the central nitrogen atom. In Fig. 1(b), a nitrogen atom is replaced by a carbon atom. Since a more electronegative nitrogen atom always carries more valence electrons than a more electropositive carbon atom, the $N_2{\rightarrow}C$ replacement minimized $|\Delta_{Max}|$ whereas $|\Delta_{Min}|$ surged to 1.0434 for a great change of valence electrons on the central carbon atom. Contrarily, the most electropositive tertiary carbon atom is preserved in $N_2{\rightarrow}C$ substitution (Fig. 1(c)) to furnish zero $|\Delta_{Min}|$ while changing a secondary nitrogen to a tertiary one renders a large $|\Delta_{Max}|$ of 1.0881. In Fig. 1(d), $|\Delta_{Std}|$ is expected to be small as both central atoms are bonded to others with the same elements regardless of their type. Nevertheless, the notable electronegativity difference between carbon and nitrogen again yields large values of $|\Delta_{Mean}|$, $|\Delta_{Max}|$, and $|\Delta_{Min}|$.

As presented in Fig. 1(a), if the $Mean$ aggregator is used alone, the message received from the central nitrogen atom's chemical environment is insufficient to differentiate the pair of molecular structures. Figs. 1(b), 1(c), and 1(d) further illustrated that using only $Max$, $Min$, and $Std$ aggregators in the cases respectively, cannot differentiate the molecular structures of the pair in each scenario. A lack of ability to differentiate structures will result in poor performance to predict structure-dependent molecule properties, such as optical band gaps. Therefore, to extract the chemical environment information effectively, it is proposed to exploit a scheme that adaptively employs multiple aggregators. These aggregators are assigned learnable weights to represent the interaction between atoms and their chemical environments. In this work, a Chemical Environment Graph Neural Network with adaptive learning is proposed to fulfill this task.

In standard GNNs, node and edge features pass through an MLP before the aggregation step. However, aggregators allow for more expressive power in capturing complex relationships and interactions between nodes in a graph. This is the intuition behind MPNNs, including GraphSAGE, GAT, GIN, EGC, and ChemGNN. Furthermore, aggregators provide a mechanism for nodes to exchange and fuse information with their neighbors. Different aggregation functions can capture different types of interactions and behaviors.

The overall framework of ChemGNN for optical band gap predictions of g-$C_3N_4$ nanosheet and its doped variants is illustrated in Fig. 8. The model takes the molecular structures as input and predicts the optical band gap (a graph-level property) of the molecules. Specifically, atoms are interpreted as nodes and chemical bonds are interpreted as edges in a graph. A stack of CEAL layers is exploited to aggregate messages from neighboring nodes to extract the underlying node embeddings. The edge features are first transformed linearly by the edge encoder to have the same size of node feature vectors. The extracted node embeddings obtained by the last CEAL layer are sent to a readout layer to form a graph-level representation. The graph representation is then used to predict the optical band gaps of g-$C_3N_4$ and its doped variants.

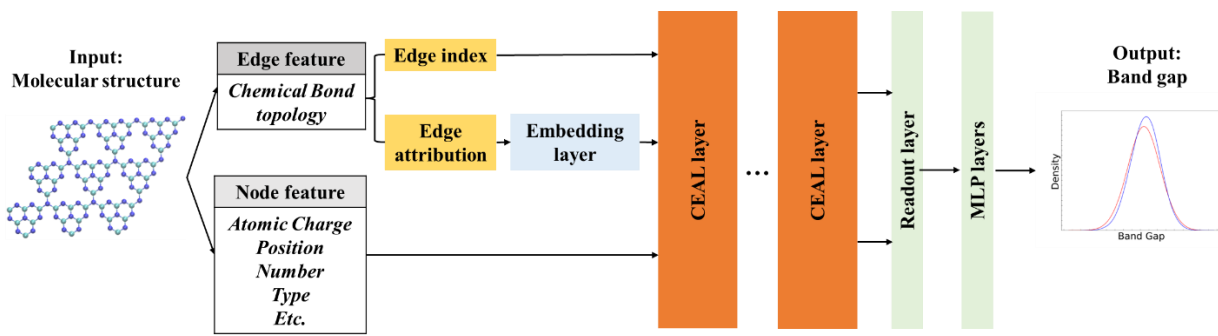

**Fig. 8. Overview of ChemGNN for optical band gap prediction of g-$C_3N_4$ nanosheet.** CEAL layers are employed to extract messages from atoms' chemical environments.

**Adaptive Aggregation Mechanism**
Our model further investigates on top of the classical GNN model PNA. In PNA, multiple aggregators are introduced. However, these aggregators have equal weights before being processed by degree-based scalers. While degree-based scalers capture the topology of graphs and have proven beneficial for networks such as social networks, they may not be sufficient for accurately learning representations of atomic interactions in chemical compounds, which are crucial for predicting molecular properties. Consider the doping operation in chemicals, where an atom is replaced by a different type of atom. For instance, the replacement of the carbon atom in CO by an oxygen atom turns the molecule into $O_2$. While the bonding topology does not change after this C→O substitution, the interatomic interactions do. To effectively learn representations of atomic interactions, we propose employing learnable weights that are applied to the aggregators. This approach allows for adaptive aggregation and enables the model to capture variations in atomic interactions, leading to improved molecular property predictions. These weights are adaptively learned in the training process to determine the optimal combination of aggregation functions based on the local chemical environments of atoms. It is achieved by assigning learnable weights to a finite collection of aggregators and formulating an arbitrary linear combination of them, written as equations (3)-(6).

$$h_i^{(k)} = \begin{bmatrix} G(w_0^{(k)}) \cdot agg\_msg_0 \\ \vdots \\ G(w_{M-1}^{(k)}) \cdot agg\_msg_{M-1} \end{bmatrix} \quad (3)$$

$$agg\_msg_A = \oplus_A \left( MLP_{msg}^{(k)}(x_i^{(k-1)}, x_j^{(k-1)}, e_{j,i}) \right), j \in N(i), A \in \{0, \cdots, M-1\} \quad (4)$$

$$G(w_A^{(k)}) = softmax(w_0^{(k)}, \ldots, w_{M-1}^{(k)}), A \in \{0, \ldots, M-1\} \quad (5)$$

where nodes $i$ and $j$ are the central and neighboring nodes, respectively. $h_i^{(k)}$ represents the message received by the central node $i$ at the $(k)th$ layer. $N(i)$ is the neighborhood of node $i$. $x_i^{(k-1)}$ and $x_j^{(k-1)}$ are the node features at the $(k-1)th$ layer. $e_{j,i}$ is the edge feature from node $j$ to $i$. $\oplus_A$ is an aggregation function. Candidate aggregators in this work are given in Table 2. $w_A^{(k)}$ is the learnable weight for aggregator $A$ at the $(k)th$ layer, $M$ is the number of aggregators. $softmax$ is used as the gate function to ensure the weight of an aggregator $G(w_A^{(k)}) \in [0,1]$ and the sum of weights $\sum_{A=0}^{M-1} G(w_A^{(k)}) = 1$. The final aggregated messages $h_i^{(k)}$ and the features of the central node $i$ are fed into the $MLP_{update}^{(k)}$ to update the embedding of the central node, which is formulated as follows:

$$x_i^{(k)} = MLP_{update}^{(k)}(x_i^{(k-1)}, h_i^{(k)}) \quad (6)$$

CEAL is a robust and compelling instantiation of the general GNN framework. $X_i^{(0)} \in \mathbb{R}^{10 \times 1}$ is the initial node feature vector, consisting of atom coordinates in space, the atom type, and the electron numbers on the 1s, 2s, 2p, 3s, 3p and 3d angular momentum channels. The architecture of a CEAL layer is illustrated in Fig. 2.

Table 2 lists the aggregation functions investigated in this work. The aggregator set $\mathcal{A} = \{Sum, Mean, Max, Min, Std\}$ contains diverse aggregate functions to extract various characteristics of an atom's chemical environment. Our candidate function set contains aggregate functions with sufficient diversity to expand the search space and improve performance. The first type of aggregators is mean aggregation $Mean_u(X^k)$ and sum aggregation $Sum_u(X^k)$, representing the average and total incoming message of node $u$ at layer $k$. $d_u = |N(u)|$ is the number of neighboring nodes. The second type of aggregators is maximum and minimum aggregations, $Max_u(X^k)$ and $Min_u(X^k)$, through which the largest and smallest neighbor incoming messages are selected. The third type of aggregators is the standard deviation or variance aggregations, $Std_u(X^k)$ or $Var_u(X^k)$, which quantify the distribution characteristics (e.g., diversity) of adjacent nodes. In addition to the above common aggregators, our CEAL layer can integrate more aggregators, such as normalized moments aggregations, $Skew_u(X^k)$, and $Kur_u(X^k)$, which are based on the $n^{th}$ root normalization and represents skewness ($n = 3$) and kurosis ($n = 4$). We expect that higher moments can better grasp the messages of neighboring nodes. The aggregation functions can be expressed as equations (7) and (8):

$$Skew_u(X^k) = \frac{\frac{1}{d_u}\sum_{v \in N(u)}\left(X_v^k - Mean_u(X^k)\right)^3}{\left(\frac{1}{d_u}\sum_{v \in N(u)}\left(X_v^k - Mean_u(X^k)\right)^2\right)^{\frac{3}{2}}} \qquad (7)$$

$$Kur_u(X^k) = \frac{\frac{1}{d_u}\sum_{v \in N(u)}\left(X_v^k - Mean_u(X^k)\right)^4}{\left(\frac{1}{d_u}\sum_{v \in N(u)}\left(X_v^k - Mean_u(X^k)\right)^2\right)^2} \qquad (8)$$

Table 2. Aggregation functions. The aggregation functions and the corresponding aggregated message for node $u$ at $kth$ layer.

| Aggregator | $A_i(\cdot)$ | Definition |
|---|---|---|
| Mean | $Mean(X) = \mathbb{E}[X]$ | $Mean_u(X^k) = \frac{1}{d_u}\sum_{v \in N(u)} X_v^k$ |
| Sum | $Sum(X) = \sum X$ | $Sum_u(X^k) = \sum_{v \in N(u)} X_v^k$ |
| Max | $Max(X) = max(X)$ | $Max_u(X^k) = \max_{v \in N(u)} X_v^k$ |
| Min | $Min(X) = min(X)$ | $Min_u(X^k) = \min_{v \in N(u)} X_v^k$ |
| Std | $Std(X) = \sqrt{\mathbb{E}[X^2] - (\mathbb{E}[X])^2}$ | $Std_u(X^k) = \sqrt{Mean_u(X^{k^2}) - Mean_u(X^k)^2}$ |

**Acknowledgments**
   **Funding:** Not applicable.

   **Author contributions:**
       Conceptualization: HC, MC, YT, YW
       Methodology: CC, EX
       Investigation: CC, EX, CY
       Visualization: CC, DY
       Supervision: HC, MC, YT, YW
       Writing—original draft: CC
       Writing—review & editing: HC, MC, YW

   **Competing interests:** Authors declare that they have no competing interests.

**Data and materials availability:** The pristine g-$C_3N_4$ nanosheet and its eight doped variants data generated in this research are provided for free research use and can be accessed from https://github.com/EnzeXu/ChemGNN_Dataset. The code for this work is publicly available at the following repository: https://github.com/EnzeXu/ChemGNN.

**Additional information:** Correspondence and requests for materials should be addressed to Hanning Chen and Minghan Chen.